\documentclass[12pt,twoside]{article}   
\usepackage[super,sort,comma]{natbib}

\usepackage{fancyhdr}		

\usepackage[section]{placeins}   %

\usepackage{graphicx}

\makeatletter \renewcommand\@biblabel[1]{$^{#1}$} \makeatother
\newcommand{\cen}[1]{\begin{center} #1 \end{center}}

\usepackage[mathlines]{lineno}

\usepackage{hyperref}
\hypersetup{ colorlinks,
    citecolor=blue,
    filecolor=blue,
    linkcolor=blue,
    urlcolor=blue
}

\usepackage{xcolor}

\definecolor{gray}{rgb}{0.6,0.6,0.6}
\definecolor{red}{rgb}{0.85,0,0}
\definecolor{green}{rgb}{0,0.85,0}
\definecolor{blue}{rgb}{0,0,0.85}
\definecolor{beige}{rgb}{0.92,0.87,0.78}
\usepackage[all]{hypcap}    

\usepackage{multirow}

\begin{document}

\cen{\sf {\Large {\bfseries Adaptive Fine-tuning based Transfer Learning for the Identification of MGMT Promoter Methylation Status} \\  
\vspace*{10mm}
Erich Schmitz$^1$, Yunhui Guo$^2$, Jing Wang$^1$} \\
$^1$Advanced Imaging and Informatics for Radiation Therapy (AIRT) and Medical Artificial Intelligence and Automation (MAIA) Laboratory, Department of Radiation Oncology, University of Texas Southwestern Medical Center, Dallas, Texas \\
$^2$Department of Computer Science, The University of Texas at Dallas
\vspace{5mm}\\
Version typeset \today\\
}

\pagenumbering{roman}
\setcounter{page}{1}
\pagestyle{plain}
Jing Wang email: jing.wang@utsouthwestern.edu \\
\begin{abstract}
\noindent {\bf Background:} Glioblastoma Multiforme (GBM) is an aggressive form of malignant brain tumor with a generally poor prognosis. Treatment usually includes a mix of surgical resection, radiation therapy, and akylating chemotherapy but, even with these intensive treatments, the 2-year survival rate is still very low. $O^{6}$-methylguanine-DNA methyltransferase (MGMT) promoter methylation has been shown to be a predictive bio-marker for resistance to chemotherapy, but it is invasive and time-consuming to determine the methylation status. Due to this, there has been effort to predict the MGMT methylation status through analyzing MRI scans using machine learning, which only requires pre-operative scans that are already part of standard-of-care for GBM patients. \\ 
{\bf Purpose:} To improve the performance of conventional transfer learning in the identification of MGMT promoter methylation status, we developed a 3D SpotTune network with adaptive fine-tuning capability. Using the pretrained weights of MedicalNet coupled with the SpotTune network, we compared its performance with an equivalent network that is initialized with random weights for different combinations of MR modalities.\\
{\bf Methods:} Using a ResNet50 as the base network, three categories of networks are created: 1) A 3D SpotTune network to process volumetric MR images, 2) a network with randomly initialized weights, and 3) a network with weights initialized from the pre-trained MedicalNet. These three networks are trained and evaluated using the UPENN-GBM dataset, a public GBM dataset provided by the University of Pennsylvania. The MRI scans from 262 patients are used, with 11 different modalities corresponding to a set of perfusion, diffusion, and structural scans. The performance is evaluated using 5-fold cross validation with a hold-out testing dataset. \\
{\bf Results:} The SpotTune network showed better performance than the network with randomly initialized weights. The best performing SpotTune model achieved an area under the Receiver Operating Characteristic curve (AUC), average precision of the precision-recall curve (AP), sensitivity, and specificity values of $0.6604$, $0.6179$, $0.6667$, and $0.6061$ respectively.\\
{\bf Conclusions:} SpotTune enables transfer learning to be adaptive to individual patients, resulting in improved performance in predicting MGMT promoter methylation status in GBM using equivalent MRI modalities as compared to using randomly initialized weights. \\

\end{abstract}

\newpage     

\tableofcontents

\newpage

\setlength{\baselineskip}{0.7cm}      

\pagenumbering{arabic}
\setcounter{page}{1}
\pagestyle{fancy}

\section{Introduction} \label{sec:intro}

Glioblastoma Multiforme (GBM) is an aggressive malignant brain tumor characterized by a generally poor prognosis and low survival rate. While it has a relatively low incidence compared to other forms of cancers, it is the most common form of primary malignant brain tumors, accounting for $45.6\%$ of them \cite{WIRSCHING2016381}. While there is one known risk factor for GBM, namely from ionizing radiation to the head, it does not account for all cases \cite{WIRSCHING2016381}. Coupling this with indeterminate onset symptoms, GBM is often diagnosed and treated late into its progression. Generally the treatment will involve surgical resection, radiation therapy, and concomitant/adjuvant temozolomide (TMZ), but even with these intensive treatments, the 2-year survival rate is only $26.5\%$ \cite{RN12}. With such a low survival, especially with treatment, a major focus GBM research is to identify clinical bio-markers that can be used to predict how a patient will respond to treatment. One potential bio-marker is the methylation status of the $O^{6}$-methylguanine-DNA methyltransferase (MGMT) promoter which provides prognostic information on how a patient will respond to TMZ \cite{RN10}.

One challenge associated with utilizing MGMT methylation as a clinical bio-marker is the complex process required to assess the methylation status. Current processes require biopsies or micro-surgical resection followed by a time-consuming molecular analysis, with no guarantees that there will be sufficient tissue available to perform a full analysis \cite{WIRSCHING2016381}. This lends to the need for a quick non-invasive process to determine methylation status. Several previous studies have explored potential in using deep learning to predict methylation status on MR imaging. Depending on the data used, there has been some success with using convolutional neural networks (CNNs) to predict the methylation status with a receiver operating characteristic (ROC) area under the curve (AUC) ranging from $58-91\%$, but due to the size of many of these study's datasets, ranging from $53-498$ patients, it is difficult to assess their generalizability\cite{RN17, RN20, RN21, RN22, RN23}. Another analysis, using the same dataset as this study, achieved an AUC of $59.8\%$ while attempting to reduce their network's parameter space through knowledge-based filtering of the MR images \cite{RN7}.

In contrast to many applications involving natural images, the sample size of medical imaging is often limited. To address this challenge and train a CNN-based model effectively, transfer learning is widely adopted. Transfer learning leverages domain knowledge from a source task to enhance the performance of a target task, a practice commonly applied when the target task has a dataset that is limited in size \cite{RN6}. Given that our classification task involves MR volumes, a related source task with valuable domain knowledge is medical image analysis/segmentation, which can be accessed through MedicalNet \cite{chen2019med3d}, a collection of weights pre-trained on a multi-class segmentation task using 3D medical images. In this study, we employed transfer learning with adaptive fine-tuning to classify MGMT promoter status, which enables evaluation tailored to individual patients.

The concept of fine-tuning is closely related to transfer learning, as it involves the optimization of the transferred network weights. In the context of this study, fine-tuning specifically refers to parameters that are allowed to adjust during training, while freezing refers to parameters that remain constant throughout training. In conventional transfer learning tasks, the choice of which parameters to fine-tune is manually determined by trial-and-error, with the most common practice being to fine-tune parameters within the last layer/s of the network. The choice of which parameters in a network to fine-tune involves a delicate balance between enabling the network to learn more about the target data by increasing the amount of fine-tuned parameters (thus risking overfitting), and preserving domain knowledge from the source task through frozen parameters (potentially not effectively learning the target dataset). To automate the choice of which parameters to fine-tune, we turn to adaptive fine tuning, eliminating the need for trial-and-error in finding the best combination of fine-tuned parameters. Due to the small size of many medical image datasets, coupled with the variability in patient characteristics, it can be difficult to extract meaningful image features in a network where all the parameters are fine-tuned. With the addition of transfer learning, image features found in the training of larger datasets are extracted, while the adaptive fine-tuning introduces a way to account for variability in image characteristics by fine-tuning on a patient-by-patient basis, allowing the network to extract the most applicable features for each image.  

In this project, we developed a SpotTune-based adaptive fine-tuning approach for methylation status prediction in GBM using MRI, where the previously developed SpotTune algorithm dynamically navigates through the fine-tuned and frozen layers within a Residual Network \cite{RN6}. We specify a Residual Network, as it has been shown to be resilient to the exchange of residual blocks since each block acts as a shallow classifier \cite{RN6}. In a SpotTune network the dynamic navigation is implemented through the exchange of residual blocks. Specific to SpotTune, the dynamic routing is determined on a per image basis, which is ideal for medical imaging due to patient variability. As part of this study, the SpotTune framework, originally developed for 2D imaging classification tasks, was extended to handle full 3D MR volumes. A key consideration in our approach is that the Residual Network used matches the network that the transfer weights are sourced from, and additionally has a straightforward implementation. To the best of our knowledge, this study represents the first use of transfer learning with adaptive fine-tuning in MGMT promoter methylation status, while other studies have only considered traditional transfer learning \cite{RN22}.  

\section{Methods and Materials}

\subsection{Dataset}\label{data}

The University of Pennsylvania glioblastoma (UPENN-GBM) cohort is a collection of 630 patients that were diagnosed with glioblastoma, and is freely available to use via The Cancer Imaging Archive \cite{RN8, RN18, upenn_dataset}. The dataset includes magnetic resonance scans with perfusion and diffusion derivatives, computational and manually derived annotations of tumor regions, radiomic features of the tumor regions, and clinical and molecular information \cite{RN8}. Included in the molecular information is the mutational status of IDH and the MGMT promoter methylation status. 

Of the 630 patients, 611 have preoperative scans comprising of four structural MRI scans: T1, post-contrast gadolinium enhanced T1 (T1-GD), T2-weighted, and T2 fluid attenuated inversion recovery (FLAIR). A subset of these patients also have diffusion tensor imaging (DTI) and dynamic susceptibility contrast (DSC) scans. The DTI scans have 4 derivative volumes: tensor's trace, axial diffusivity, readial diffusivity and fractional anisotropy \cite{RN8}. The DSC scans have 3 derivative volumes: peak height, percentage signal recovery, and an automated proxy to the relative cerebral blood volumes \cite{RN8}. There are 291 patients that have the MGMT promoter methylation status available, of these there are 262 who have the corresponding pre-operative scans with 151 not methylated and 111 methylated. The provided images used were already converted from the DICOM format into the Neuroimaging Informatics Technology Initiative (NIfTI) format, following the processing protocol of the Brain Tumor Segmentation (BraTS) challenge \cite{RN8}. This preprocessing included de-identification, de-facing, re-orientation of images to the left-posterior superior coordinate system, registration and resampling to an isotropic resolution of 1 $mm^2$ based on the SRI common anatomical atlas, and an N4 bias field correction \cite{RN8}. Additionally segmentation masks are provided for 3 tumor subregions, the enhancing tumor, necrotic tumor core and the edematous tissue. These masks were automatically generated with some manual refinement where needed.

From the full cohort, 240 patients and their imaging were selected for this study. These correspond to the patients that have a pre-operative base scan and the MGMT methylation status available, minus 22 that were dropped due to how the training datasets between different modalities were split. For the structural imaging modalities the 240 patients are used, while for the DTI and DSC volumes, there are 208 and 189 patients respectively. Of the patients with a DSC perfusion scan, 108 are not methylated and 81 are methylated, and for those with DTI scans 127 are not methylated and 91 are methylated.

\subsection{Data Preparation}\label{prep}

Using the NIfTI files provided in the UPENN-GBM dataset, further preprocessing was applied to the images before they were used for training and testing. The images and masks were read in using the SimpleITK software package \cite{RN19}, and converted into numpy SpotTune, Randoms. The image masks were applied, and the size of the images were reduced by cropping following the largest tumor among the patients, and down-scaling the images by a factor of 2. This reduced the dimensions of the volumes respectively from $155\times 240\times 240 \rightarrow 140\times 172\times 164 \rightarrow 70\times 86\times 82$. For the final image size, the last dimension was padded to have a dimension of $70\times 86\times 86$ with a $2$ $mm$ spacing. Lastly the images had a min-max scaling applied separately for each patient, with a resulting image range of $[0, 1]$. 

The dataset was split into a training and testing set at a 2:1 ratio, and the training dataset further split using stratified 5-fold cross-validation (CV), with four of the folds being used for training and the fifth for validation. This came out to having a training set of 183, 151, and 132 patients for the structural, DTI, and DSC modalities respectively and a common testing set of 57 patients. The datasets were split such that the different MR modalities would share patients in the training and testing sets, i.e., the patients that are in the testing set for the DSC derivatives will be the same set of patients contained in the testing set of the structural modalities. So that the results are directly comparable between modalities, only the patients common to the DSC modalities were retained in the testing dataset for the structural and DTI modalities. After the training set was split into its 5 folds, each set of 4 folds used for training were augmented in four ways including flipping the images, randomly rotating the images between $-180$ and $180$ degrees, adding Gaussian noise, and applying an elastic deformation \cite{gijs_van_tulder_2022_7102577}. To account for the class imbalance of the dataset, the augmented training folds were randomly pruned until the ratio of the MGMT classes were equal. The class imbalance was kept in the validation and testing datasets.

\subsection{Model}

The 3D SpotTune network developed for this study is an amalgamation of multiple residual networks (ResNets) that allows for a dynamic routing through sets of fine-tuned and frozen pre-trained residual blocks. As SpotTune was initially produced to run on 2D images, it was further developed to run on 3D images. The bulk of the conversion to a 3D network involved changing out the 2D convolutional layers with their 3D counterparts. Additionally, the padding of the layers needed to be adjusted for the shape of the 3D inputs, due to the differences in the dimensions. The rest of the adjustments for the 3D network were done during the network training, involving the optimization of hyper-parameters specific to a SpotTune network, detailed in Section \ref{sec:network}. The optimization of these hyper-parameters ended up being the most challenging part of implementing the 3D network, as the 3D SpotTune network has $40.5$ million more trainable parameters than its 2D counterpart, increasing from $19.5$ to $60$ million.

The SpotTune network consists of three separate ResNets, one agent and two base networks. The base neural network used is a ResNet50 \cite{RN11}, which consists of 16 residual blocks adding up to 50 layers in total. Two of these networks make up the main network, where one ResNet50 contains a set of frozen residual blocks, while the other contains fine-tuned blocks. The agent network is a ResNet10 and is used to make the policy that the main network follows in how to route the data through the frozen and fine-tuned blocks. With 16 residual blocks for the ResNet50, the agent network has 16 classes as output, with 2 categories each corresponding to whether a block is to be frozen or fine-tuned. The routing is determined following Equation \ref{eq:routing}, describing the $\ell$-th residual block, where $x$ is the input image, $F_{\ell}$ is the $\ell$-th block from the frozen ResNet50, $\hat{F}_\ell$ is the $\ell$-th block from the fine-tuned ResNet50 and a duplicate of $F_{\ell}$, and $I_{\ell}(x)$ is a binary variable taken from the policy output by the agent network \cite{RN6}. 

\begin{equation}
	\label{eq:routing}
	x_\ell = I_{\ell}(x)\hat{F}_{\ell}(x_{\ell-1}) + (1-I_{\ell}(x))F_{\ell}(x_{\ell-1}) + x_{\ell-1}
\end{equation}

\begin{figure}
	\centering
	\includegraphics[width=180mm]{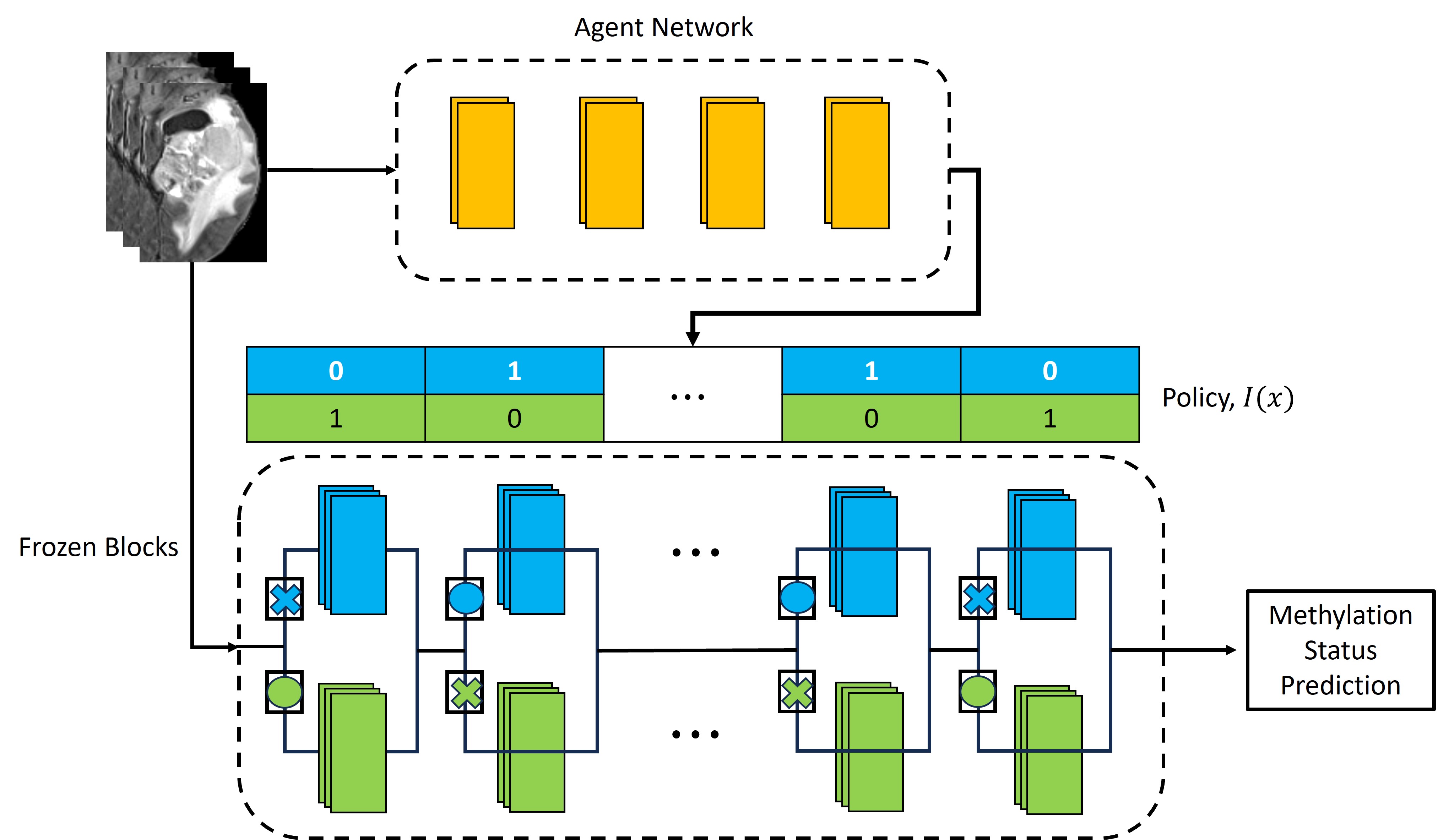}
	\caption{A diagram illustrating the SpotTune Network. The Agent network produces the policy used by the main network. The main network takes this policy, shown as a vector of binary values, and routes inputs through the residual blocks of the main network, with the routing illustrated using X's and O's.}
	\label{fig:spottune}
\end{figure}

The policy output by the agent network is a collection of binary variables that is discrete and non differentiable. In order to allow back-propagation through the full network a Gumbel Softmax sampling approach is used \cite{RN6}. In the forward pass of the network a Gumbel distribution parameterized by the logits of the agent network is used to produce a discrete sample following Equation \ref{eq:arg_max}, where $\alpha_i$ corresponds to the output of the policy network, $G_i$ is the standard Gumbel distribution, and $X$ corresponds to a discrete sampling of the policy, $I_l(x)$ \cite{RN6}. In the backward pass instead of using Equation \ref{eq:arg_max}, the Gumbel Softmax distribution is used as a relaxation of Equation \ref{eq:arg_max}, and is given in Equation \ref{eq:g_soft}, where $\tau$ is the temperature parameter that controls the discreteness of the distribution \cite{RN6}. By sampling from this distribution, the discrete policy is relaxed to be continuous and differentiable, allowing back propagation of the policy through the agent network. In other words, $Y$ is a continuous form of $X$, which is a discrete sampling of the policy, $I(x)$, where $Y$ is used to ensure that the gradient of the loss function from the main network can be backpropagated through to the agent network. Figure \ref{fig:spottune} illustrates the SpotTune network and how it would route inputs through the main network based on a policy given by the agent network.

\begin{equation}
	\label{eq:arg_max}
	X = \arg \max [\log \alpha_i + G_i]
\end{equation}

\begin{equation}
	\label{eq:g_soft}
	Y_i = \frac{exp((\log \alpha_i + G_i)/\tau)}{\sum_{j=1}{2}exp((\log \alpha_j + G_j)/\tau)}
\end{equation}

When the agent network creates a routing policy, it does so for each image, $x$, provided as input. This means that each image gets a unique policy, $I(x)$, that determines how blocks should be swapped in the main network. In the forward pass of training, the agent network will give a discrete policy for each input, that the main network will then follow, swapping between the frozen and fine-tuned residual blocks as designated. In the backward pass the gradient from the main network will be back propagated through the agent network allowing it to refine the policy. The agent network is then jointly trained with the main network to determine the optimal fine-tuning strategy and thus maximize the performance metric \cite{RN6}. Additionally, with the proposed adaptive fine-tuning, depending on the input images, each patient will have a unique route through the fine-tuned blocks of the pre-trained model, which can lead to a more personalized prediction.

In addition to the SpotTune network, a regular ResNet50 is used for comparison purposes. The ResNet50 is initialized with two different sets of weights, corresponding to random weight initialization following a uniform distribution and the transferred MedicalNet weights as used in the SpotTune network. The ResNet50 with transferred weights is initialized so that only the classification layer is fine-tuned, keeping all other layers frozen. 

\subsection{Network Training and Evaluation} \label{sec:network}

The networks were trained using the pytorch framework\cite{NEURIPS2019_9015} with a maximum of 90 epochs, a batch size of 12, and a starting learning rate of 1e-5 for the main network and 1e-4 for the agent network. The learning rate followed a decrease-on-plateau based scheduler with a patience of 20 epochs and a multiplicative factor of 0.1. Additionally, the temperature, $\tau$, of the Gumbel Softmax distribution was set to 100. The objective function used in the network was binary cross-entropy loss, with the Adam algorithm used for optimization. The inputs to the network are single channel volumes corresponding to each available modality/derivative for a total of 11 modalities: 4 structural, 4 DTI, and 3 DSC. Before deciding on training with a single channel, we also looked at multi-channel inputs ranging from 3 up to 11 channels, with each channel corresponding to a modality/derivative. In order to implement multiple channels the MedicalNet weights had to be adjusted since they assume single channel inputs. To accomplish this adjustment, we followed a weight inflation method that duplicates weights along a desired axis, and averages them over the number of duplications \cite{pmlr-v193-zhang22a}. For example, for a 3-channel input, the weights of the first convolutional layer are duplicated 3 times along the channel axis, with those weights averaged over the 3 channels. The network was trained using 5-fold cross-validation and evaluated using the accuracy (ACC), area under the receiver operating characteristics curve (AUC), sensitivity (SEN), and specificity (SPE) metrics from the torchmetrics package \cite{Detlefsen2022}. The best models to be used for the hold-out testing dataset were determined by evaluating the performance of the validation folds using an evaluation criterion, $M$,  based on the sensitivity and specificity, and given in Equation \ref{eq:m_metric} \cite{RN13}. 

\begin{equation}
	M = 0.6\times (SEN \geq 0.5) + 0.4\times (SPE \geq 0.5)
	\label{eq:m_metric}
\end{equation} 

As part of the training process the hyper-parameters were tuned using a random grid search to get an initial parameter set. With this set as basis, further independent scans were performed on a selection of hyper-parameters. These independent scans included the number of epochs with a range of $30-200$, learning rate with a range of $0.1-1e-8$, and Gumbel Softmax temperature, with a range of $1e-4-1e4$. The metric used for validation was initially a combination of ACC and AUC, but discovered that this metric selected models with a severely unbalanced sensitivity and specificity. To rectify this, we changed to a metric that prefers a balanced sensitivity and specificity, as seen in Equation \ref{eq:m_metric}, following what was done in another study\cite{RN13}. As mentioned previously, a major concern of training with 3D inputs is the sensitivity to overfitting due to the large number of parameters, so as part of the hyper-parameter scan we looked at the degree of overfitting that was occurring on the training dataset. 
 
The models chosen for evaluation on the testing dataset were combined into various sets of ensembles based on their imaging modalities and network used. Due to the random sampling inherent to the Gumbel Softmax distribution, each model created using the SpotTune network was evaluated 100 times using a random seed of 42. These 100 samples were combined to form the chosen model for each of the folds in order to better encompass the distribution of probabilities possible from the random sampling. The chosen models for the CV folds were combined by taking the mean of the output probabilities to get a single model for each imaging modality, and then the models for derivatives of each modality were combined in the same way to get overall models. The models presented in the results fall under 3 categories: those using the SpotTune network, those using randomly initialized weights in a regular ResNet50 and those that use the transferred weights in a regular ResNet50. For each of these categories, the overall models that were evaluated are labeled as DSC, DTI, Structural, and then combinations of the three. The DSC and DTI models are ensembles of their derivative volumes, and the Structural model is an ensemble of the four structural modalities T2, FLAIR, T1, and T1GD. The additional modality combinations correspond to DSC and DTI (DSC+DTI), DSC and Structural (DSC+Struct), DTI and Structural (DTI+Struct), and the three together (DSC+DTI+Struct). 

In addition to the deep learning models from this study, we compared our results with another model that achieved superior performance (AUC = $0.88$ on a private dataset) in predicting MGMT promoter methylation status using a set of 6 radiomic features in a random forest classifier \cite{RN23}. Rather than directly comparing the published results with ours, we applied their model on our dataset, due to the difference in datasets. Following the published model, we calculated the 6 radiomic features that were used: (1) Skewness from the T1 core, (2) energy from the T1 edema invasion, (3) GLCM contrast from the FLAIR necrotic core, (4) GLSZM gray level variance from the T1GD enhanced area, (5) GLSZM low gray level zone emphasis from the T2 edema invasion, and (6) NGTDM busyness from the T2 core. Each of these features came from different modalities and masks, where the core refers to the combination of the enhanced area and the necrotic core \cite{RN23}. The features were used as input to a random forest classifier and trained with the same dataset splitting as this study. The optimal hyperparameters of the random forest were set as following: a number of estimators of $10,000$, a maximum tree depth of $5$, and a minimum number of samples at a leaf node of $10$.


\section{Results}

We compare ensemble models for three different network categories. The main comparisons are between models using the same inputs, only considering cross modality comparisons for the top models for each network category. Tables \ref{tab:spottune}, \ref{tab:random}, and \ref{tab:transfer} give the results for the SpotTune, random weight initialization, and transferred weight initialization respectively. Included in the tables are the AUC of ROC curves and the Average Precision of precision-recall curves (AP), with the AP given to show how well the models classify the positive, i.e. methylated, samples. Additionally the sensitivity and specificity are included to better show any imbalance in how the models classify the positive and negative samples.

\begin{table}[h!]
	\centering 
	\begin{tabular}{c c c c c}
		\hline
		\multicolumn{5}{c}{SpotTune Results} \\
		\hline
		Model & AUC & AP & SEN & SPE \\
		\hline
		DSC & 0.6124 & 0.5842 & 0.5000 & 0.6061 \\
		DTI & \textbf{0.6654} & 0.6008 & 0.7500 & 0.6364 \\
		Struct & 0.5707 & 0.5346 & 0.5833 & 0.4848 \\
		DSC+DTI & 0.6604 & \textbf{0.6179} & 0.6667 & 0.6061 \\
		DSC+Struct & 0.6073 & 0.5962 & 0.5417 & 0.5455 \\
		DTI+Struct & 0.6376 & 0.5289 & 0.5833 & 0.6364 \\
		DSC+DTI+Struct & 0.6389 & 0.5779 & 0.6667 & 0.5758 \\
		\hline	
	\end{tabular}
	\caption{The ROC AUC, average precision, sensitivity and specificity evaluated on the hold out testing dataset using model ensembles created with the SpotTune network. The highest values for the ROC AUC and AP are given in bold text.}
	\label{tab:spottune}
\end{table}

\begin{table}[h!]
	\centering
	\begin{tabular}{c c c c c}
		\hline
		\multicolumn{5}{c}{Randomly Initialized Weights Results} \\
		\hline
		Model & AUC & AP & SEN & SPE \\
		\hline	
		DSC & 0.5051 & 0.4872 & 0.3750 & 0.5758 \\
		DTI & 0.6023 & 0.5159 & 0.3750 & 0.6364 \\
		Struct & \textbf{0.6061} & \textbf{0.5334} & 0.4167 & 0.6061 \\
		DSC+DTI & 0.5707 & 0.5127 & 0.3333 & 0.6364 \\
		DSC+Struct & 0.5669 & 0.5271 & 0.3750 & 0.6667 \\
		DTI+Struct  & 0.5997 & 0.5187 & 0.4167 & 0.6667 \\
		DSC+DTI+Struct & 0.5783 & 0.5018 & 0.4583 & 0.6364 \\
		\hline
	\end{tabular}
	\caption{The ROC AUC, average precision, sensitivity and specificity evaluated on the hold out testing dataset using model ensembles created with a ResNet50 initialized with random weights. The highest values for the ROC AUC and AP are given in bold text.}
	\label{tab:random}
\end{table}

\begin{table}[h!]
	\centering
	\begin{tabular}{c c c c c}
		\hline
		\multicolumn{5}{c}{Transferred Weights Results} \\
		\hline
		Model & AUC & AP & SEN & SPE \\
		\hline
		DSC & 0.5303 & 0.4421 & 0.5000 & 0.6970 \\
		DTI & \textbf{0.5947} & \textbf{0.5445} & 0.5833 & 0.4848 \\
		Struct & 0.4722 & 0.4250 & 0.5417 & 0.4242 \\
		DSC+DTI & 0.5808 & 0.4992 & 0.5417 & 0.5455 \\
		DSC+Struct & 0.5051 & 0.4296 & 0.5000 & 0.5455 \\
		DTI+Struct & 0.5732 & 0.4869 & 0.6250 & 0.6061 \\
		DSC+DTI+Struct & 0.5682 & 0.4788 & 0.5000 & 0.6061 \\
		\hline	
	\end{tabular}
	\caption{The ROC AUC, average precision, sensitivity and specificity evaluated on the hold out testing dataset using model ensembles created with a ResNet50 initialized with transferred MedicalNet weights, while only fine-tuning the classification layer. The highest values for the ROC AUC and AP are given in bold text.}
	\label{tab:transfer}
\end{table}

\begin{figure}
	\centering
	\includegraphics[width=80mm]{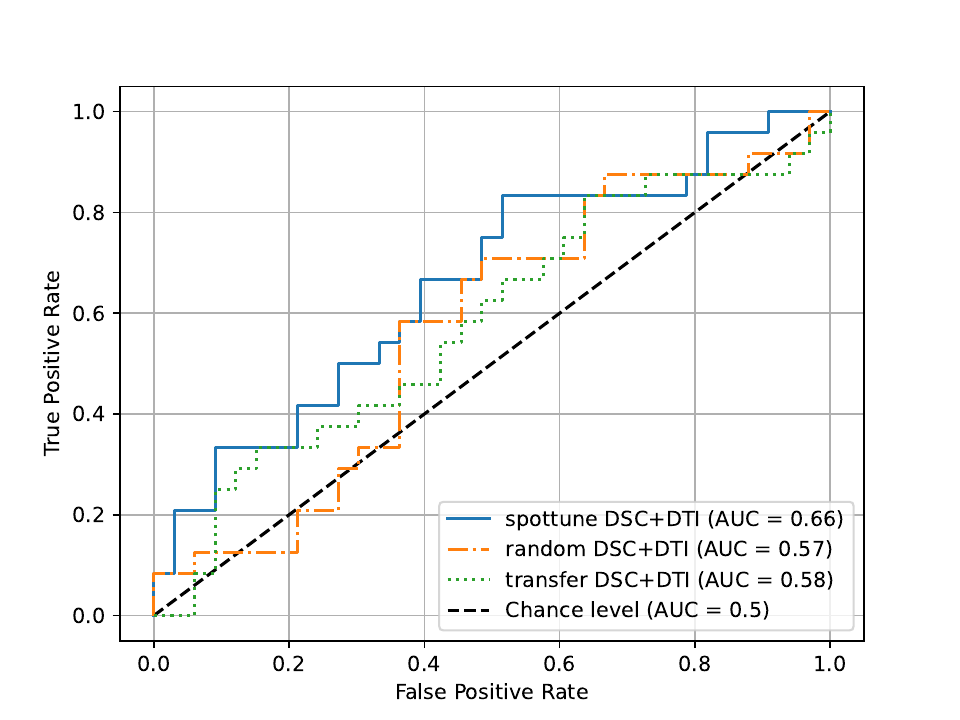}
	\caption{ROC curves for the DSC+DTI model ensembles for the three network categories. }
	\mbox{}
	\mbox{}
	\label{fig:roc_dsc_dti}
\end{figure}

\begin{figure}
	\centering
	\includegraphics[width=80mm]{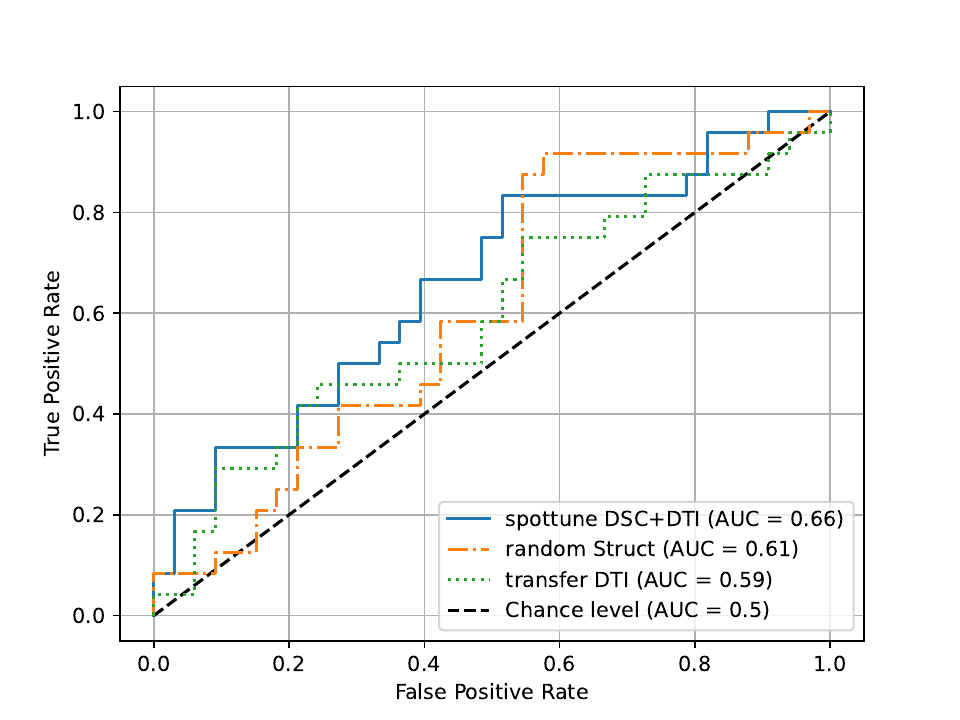}
	\caption{ROC curves for the best overall model ensembles for the three network categories.}
	\mbox{}
	\mbox{}
	\label{fig:roc_overall}
\end{figure}

\begin{figure}
	\centering
	\includegraphics[width=80mm]{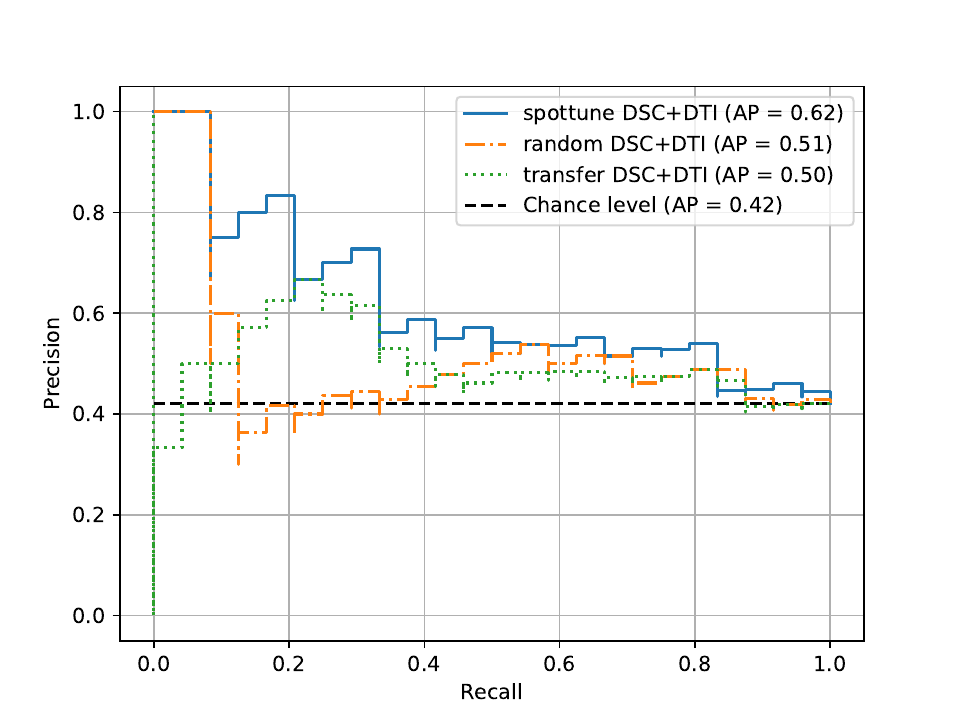}
	\caption{Precision-recall curves for the DSC+DTI model ensembles for the three network categories. }
	\mbox{}
	\mbox{}
	\label{fig:pr_dsc_dti}
\end{figure}

\begin{figure}
	\centering
	\includegraphics[width=80mm]{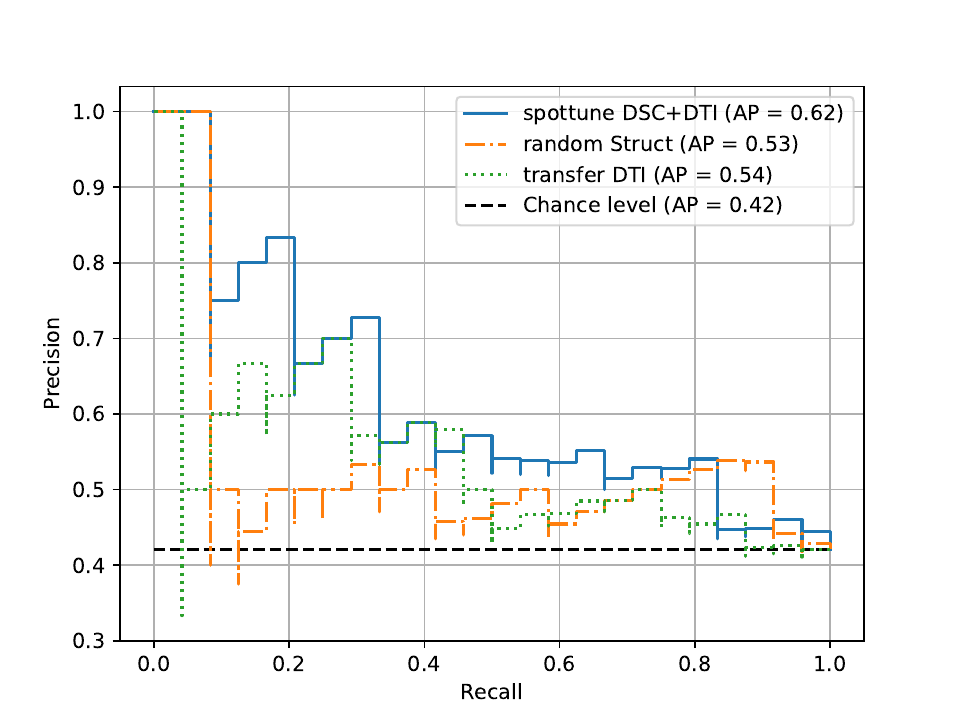}
	\caption{Precision-recall curves for the best overall model ensembles for the three network categories.}
	\mbox{}
	\mbox{}
	\label{fig:pr_overall}
\end{figure}

With the exception of the model using the structural modalities on their own, the SpotTune network outperforms the randomly initialized network for similar inputs. The highest overall performance is given by the DSC+DTI SpotTune model with a ROC AUC of $0.6604$, and an AP of $0.6179$. These results outperform the corresponding DSC+DTI model using the ResNet50 with randomly initialized weights, which had a ROC AUC of $0.5707$ and an AP of $0.5334$. The same DSC+DTI model using transferred weights with a ResNet50 showed a slightly better performance when only looking at the ROC AUC of $0.5808$, but the picture changes when also considering the AP of $0.4992$, showing the the non-adaptive transfer learning is not sufficient on its own. ROC Curves of the DSC+DTI models for each of the network categories can be found in Figure \ref{fig:roc_dsc_dti}, and curves for the overall best models for the three categories is found in Figure \ref{fig:roc_overall}. Additionally the precision-recall curves are given for the same ensembles in Figures \ref{fig:pr_dsc_dti} and \ref{fig:pr_overall}.

In addition to the improved overall performance of the SpotTune models for similar inputs, there is a noteworthy observation with regards to the sensitivity and specificity. The SpotTune models have a better balance and higher sensitivity than the randomly initialized models in all of the ensembles. The sensitivity for the randomly initialized model tends to be low, with a $SEN/SPE$ ratio as low as $0.5$, with a maximum sensitivity between the models of $0.4583$, compared to the maximum in the SpotTune models of $0.7500$. Applying traditional transfer learning as shown in Table \ref{tab:transfer} improves the sensitivity/specificity balance over the randomly initialized weights and the addition of the SpotTune network further improves it. 

For the comparison radiomics model on the dataset used in this study, the random forest classifier with 6 chosen radiomic features achieved a ROC AUC of $0.5764$, an AP $0.4886$, a sensitivity of $0.2174$ and a specificity of $0.9355$. Since this comparison model only made use of the structural modalities we can compare it to the SpotTune model that used the structural modalities as input. The ROC AUC of the two models are similar at $0.5707$ and $0.5764$, but the SpotTune gave a better AP of $0.5346$ compared to the $0.4886$ of the comparison radiomics model. For the best overall SpotTune model, both the ROC AUC and AP showed an improved performance. Additionally the SpotTune models had a better balance between the sensitivity and specificity, compared to the high imbalance for the comparison radiomics model.

The significance, $p<0.05$, of the results was checked by performing a Friedman chi-square test followed by a post-hoc Nemenyi test should the Friedman test prove significant. The p-values for the two sets tests are given in Table \ref{tab:p_values_nemenyi},  and compare the models that share the same input between the SpotTune, non-adaptive transfer learning, and randomly initialized networks. From the Friedman test, four sets of models were significant with $p<0.05$: The DSC+DTI, DSC+Struct, DTI+Struct and DSC+DTI+Struct. Of these four sets of models, four specific comparisons showed a significant difference in the results following the Nemenyi test. Three of the comparisons corresponded to the SpotTune and randomly initialized networks, with the last between the non-adaptive transfer learning and randomly initialized networks. It is noteworthy that each SpotTune network that showed significance involved the DTI modalities as part of their input, but only when coupled with other modalities. If we loosen the significance to $p<0.1$ the best overall models then pass the Friedman test, with each set of comparisons showing significance using the Nemenyi test.

\begin{table}[h!]
	\centering
	\begin{tabular}{c c c c}
		\hline
		\multicolumn{4}{c}{Friedman chi-square \& Nemenyi p-values} \\
		\hline
		Model & Friedman p-value & Networks Compared & Nemenyi p-value \\
		\hline
		\multirow{3}{*}{DSC} & \multirow{3}{*}{0.108} & SpotTune, Random & 0.177\\
		&& Random, Transfer & 0.147 \\
		&& SpotTune, Transfer & 0.9 \\
		\hline
		\multirow{3}{*}{DTI} & \multirow{3}{*}{0.128} & SpotTune, Random & 0.121 \\
		&& Random, Transfer & 0.339 \\
		&& SpotTune, Transfer & 0.822 \\
		\hline
		\multirow{3}{*}{Struct} & \multirow{3}{*}{0.0970} & SpotTune, Random & 0.0982 \\
		&& Random, Transfer & 0.249 \\
		&& SpotTune, Transfer & 0.876 \\
		\hline
		\multirow{3}{*}{DSC+DTI} & \multirow{3}{*}{\textbf{0.0131}} & SpotTune, Random & \textbf{0.0103}\\
		&& Random, Transfer & 0.147\\
		&& SpotTune, Transfer & 0.554\\
		\hline
		\multirow{3}{*}{DSC+Struct} & \multirow{3}{*}{\textbf{0.0222}} & SpotTune, Random & 0.0503\\
		&& Random, Transfer & \textbf{0.0395}\\
		&& SpotTune, Transfer & 0.9 \\
		\hline
		\multirow{3}{*}{DTI+Struct} & \multirow{3}{*}{\textbf{0.00606}} & SpotTune, Random & \textbf{0.00567}\\
		&& Random, Transfer & 0.0635 \\
		&& SpotTune, Transfer & 0.661 \\
		\hline
		\multirow{3}{*}{DSC+DTI+Struct} & \multirow{3}{*}{\textbf{0.0180}} & SpotTune, Random & \textbf{0.0181}\\
		&& Random, Transfer & 0.0982\\
		&& SpotTune, Transfer & 0.768\\
		\hline
		\multirow{3}{*}{Best Overall} & \multirow{3}{*}{0.0636} & SpotTune, Random & 0.0181\\
		&& Random, Transfer & 0.0982\\
		&& SpotTune, Transfer & 0.0768\\
		\hline	
	\end{tabular}
	\caption{p-values calculated from the Friedman chisquare and nemenyi tests for comparing the spottune, randomly initialized and non-adaptive transfer learning networks. Comparisons are between models that used the same inputs, and then one for the best overall model from each network. The best overall models compared are the DSC+DTI, Struct, and DTI for the SpotTune, randomly initialized and non-adaptive transfer learning networks, respectively. Comparisons that show significance, $p<0.05$, are shown in bold}
	\label{tab:p_values_nemenyi}
\end{table}

\section{Discussion And Conclusion}

In this study we examine how transfer learning with adaptive fine-tuning has the potential to improve the prediction of MGMT methylation status over the use of randomly initialized weights in similar networks. Using the SpotTune network, we showed improved results compared to randomly initialized counterparts for various model ensembles. It is well known that the volume of data in a machine learning experiment can greatly affect performance, and dataset size is an especially important concern with relation to medical imaging \cite{chen2019med3d}. Through the use of transfer learning, the lower level information that is gained from a large dataset can be applied to a smaller dataset with varying degrees of fine-tuning. With transfer learning specifically, a concern in medical imaging is the availability of weights trained on a sufficiently large dataset that also contains necessary domain knowledge. This concern is partially abrogated through the MedicalNet weights, which are trained on 3D segmentation tasks using MRI and CT scans from varying medical datasets. While the dataset is not as large as the ImageNet dataset, which is more commonly used in transfer learning tasks, it contains 3D domain knowledge more specific to the target task. The addition of adaptive fine-tuning to the transfer learning works to remove a layer of complexity in the training, namely optimization of which layers to fine-tune and freeze. This difference between traditional fine-tuning and adaptive fine-tuning is seen in the improvement of results in Table \ref{tab:spottune} over Table \ref{tab:transfer}, not including the reduction in the usage of computational resources since there was no need to run dozens of trainings to optimize the selection of fine-tuned layers.   

We compared our results to two different studies, one which used the same dataset \cite{RN7}, and another in which we attempt to reproduce their results on the dataset used in this study \cite{RN23}. The first study reported a ROC AUC of $0.5980$, a sensitivity of $0.4535$ and a specificity of $0.7403$\cite{RN7}, and the model used by the second study and applied to our dataset achieved a ROC AUC of $0.5764$, an AP of $0.4886$, a sensitivity of $0.2174$ and a specificity of $0.9355$. When comparing the results gained from the same dataset, our best overall model with a ROC AUC of $0.6604$ and an AP of $0.6179$ was able to produce both an improved ROC AUC and AP, and additionally a better balance between the sensitivity and specificity. The second study performed their analysis on a private dataset and reported a ROC AUC of $0.88$, a sensitivity of $0.70$ and a specificity of $0.86$. This is a stark difference to what we were able to achieve when duplicating the model on the public dataset used in this work and provides some insight into whether reported results on private datasets can be directly compared. While our results do not necessarily perform as well as those studies \cite{RN17, RN20, RN21, RN22, RN23} mentioned in the introduction section, especially those who report a ROC AUC $>0.8$ \cite{RN20, RN21, RN22, RN23}, it is important to note that the differences in datasets make direct comparisons difficult, where there is a need to test models on equivalent datasets before making proper comparisons.  

With regards to optimizing the fine-tuned layers in traditional transfer learning, it may be noticed that there is one outlier in Table \ref{tab:transfer}, where the Struct model shows a performance less than $0.5$ for the AUC. One potential reason for this is that the lower level features selected by the transferred weights do not correlate as well with the structural modalities, since the classification layer is the only one that is fine-tuned. Due to this, there were a number of training folds that were not able to produce a model that passes the selection criteria, forcing the training to use the model from the last epoch. For the Struct model, there were 5 training folds that were not able to produce models that passed the selection criteria, whereas this number was only 3 and 1 for the DSC and DTI models, respectively. To address this a second round of training was done for the structural modalities using the ResNet50 with transferred weights that relaxed the number of frozen layers, allowing the last two residual blocks, in addition to the classification layer, to be fine-tuned. The resulting model gave a more reasonable performance with an AUC of $0.5341$, an AP of $0.5134$, a SEN of $0.5417$, and a SPE of $0.4848$. These results support the idea that the strictness of the fine-tuning posed challenges for the Struct model, and that the transferred weights used were not as useful for these modalities.

In addition to the use of adaptive fine-tuning for improved results, the choice of imaging modalities also plays an important role in performance. In this study, each ensemble model was made by combining models trained on separate imaging modalities made available in the UPENN-GBM dataset. Important decisions included whether these different modalities should be included as separate channels in a single training, and which combination of models, if any, would yield optimal results. When considering the number of channels, it was found that keeping the trainings to a single channel was optimal, especially with regards to the MedicalNet weights, which assumes a single channel. This can be seen in a separate training done using the 3 DSC derivatives as 3 channels in a single training, resulting in an AUC of $0.5758$, compared to the AUC of $0.6124$ given in Table \ref{tab:spottune} for the DSC model, which is an ensemble of the 3 DSC derivative models. For the 3-channel case the weights were generated using a technique known as weight inflation \cite{pmlr-v193-zhang22a}, where the single channel weights were duplicated and normalized for 3 channels. An interesting outcome related to the choice of imaging modality was the improved performance of models that used the diffusion and perfusion modalities over the more common structural modalities. 

The main limitation of this work is the size of the dataset, with only 262 patients having the MGMT methylation status labeled. While transfer learning was used as a way to counteract this, it still proved an issue, especially with regards to overfitting. To increase the size of the dataset, it would be possible to move from prediction of MGMT methylation status to one of treatment response in a future study. Such a course would have the potential of doubling the size of the dataset with respect to this study, since the overall dataset has 630 patients. Another limitation is the combinatorics problem related to how the modalities should be combined to create on overall model. Apart from the choice of the number of channels to include in the input and how the models from different modalities should be combined, there is the question of which combinations of modalities to use as the channels or model ensembles. This study took a more conservative approach and focused on keeping the like modalities together in an ensemble, i.e. keeping the DSC, DTI and structural modalities together, where there is the option of using cross combinations of modalities. One example could be creating an ensemble of the pulse height derivative of the DSC modality with the tensor's trace derivative of the DTI modality.

\section{Acknowledgment}

We acknowledge the funding support from the National Institutes of Health (R01CA251792).

\section{Conflict of Interest}

The authors have no relevant conflicts of interest to disclose.

\section{Data Availability}

The UPENN-GBM dataset is publicly available through the Cancer Imaging Archive \cite{RN8, RN18, upenn_dataset}. The code used for this study is publicly available on GitHub at \url{https://github.com/LabAIRT/gbm_project}

\clearpage


\section*{References}
\addcontentsline{toc}{section}{\numberline{}References}
\vspace*{-20mm}









\bibliographystyle{./medphy.bst}    


\end{document}